\documentclass[preprint,prd,amsmath,amssymb,floatfix,nofootinbib]{revtex4}

\usepackage[colorlinks=true,urlcolor=black]{hyperref}

\hypersetup{backref = true,pagebackref = true,hyperindex = true, colorlinks = true,
  breaklinks = true, urlcolor = blue, linkcolor = blue, bookmarks = true,
  bookmarksopen = true, citecolor=red}

\usepackage{bm}
\usepackage{color}
\usepackage{dcolumn}
\usepackage{graphics}
\usepackage{graphicx}
\usepackage{subfigure}
\usepackage{slashed}
\usepackage{placeins}

\allowdisplaybreaks
\usepackage{pdfpages}
\usepackage{eso-pic}
\usepackage{everyshi}
\usepackage{pdflscape}

\def\al{\alpha}

\def\veps{\varepsilon}

\newcommand{\ep}{\varepsilon}

\def\be{\begin{equation}}
\def\ee{\end{equation}}
\def\bea{\begin{eqnarray}}
\def\eea{\end{eqnarray}}
\def\bse{\begin{subequations}}
\def\ese{\end{subequations}}
\def\bc{\begin{center}}
\def\ec{\end{center}}

\def\nonum{\nonumber}

\def\I{{\rm i}}

\def\D{{\rm d}}

\newcommand{\eg}{{\it e.g.}}

\begin{document}

\title{
On  Landau-Khalatnikov-Fradkin transformation
in quenched QED$_3$.
}
\author{
  A.\ V.~Kotikov
       }
       \affiliation{
         Bogoliubov Laboratory of Theoretical Physics, Joint Institute for Nuclear Research, 141980 Dubna, Russia.
 }

\date{\today}

\begin{abstract}
We present the results of studies \cite{Gusynin:2020cra,Pikelner:2020mga} of the gauge covariance of the massless fermion propagator in three-dimensional
quenched quantum electrodynamics in the framework of dimensional regularization in $d=3-2\ep$.
Assuming the finiteness of the perturbative expansion, i.e. existence of the limit $\ep \to 0$, it was shown in \cite{Gusynin:2020cra} that exactly
for $d=3$ all odd perturbative coefficients, starting from the third order, must be equal to zero in any gauge.
To test this, in Ref. \cite{Pikelner:2020mga} we calculated three- and four-loop corrections to the massless fermionic propagator.
Three-loop corrections are finite and gauge-invariant, while four-loop corrections have singularities.
The terms depending on the gauge parameter are completely determined by the lower orders in accordance with the Landau-Khalatnikov-Fradkin transformation.
\end{abstract}

\maketitle

\section{Introduction}
\label{Sec:Introduction}

Quantum electrodynamics in three space-time dimensions of  (QED$_3$) with $N$ flavors of four-component massless Dirac
fermions has been under continuous study for the past forty years as a useful field-theoretical model.
QED$_3$ served as a toy model for exploring several key problems in quantum field theory, such as infrared singularities in
low-dimensional massless particle theories, coupling constant nonanalyticity in perturbation theory, dynamic symmetry breaking
and fermion mass generation, phase transition and the relationship between chiral symmetry breaking and confinement.

Moreover,
QED$_3$ has found many applications also in condensed matter physics, in particular,
in superconductivity with high $T_c$ \cite{Dorey1992}, in planar antiferromagnets \cite{Farakos1998}, as well as in studies of
graphene \cite{Semenoff1984}, where excitations of quasiparticles have linear dispersion at low energies and are described by the
massless Dirac equation in $2+1$ dimensions (see reviews on graphene studies in \cite{reviews}).

Massless QED$_3$ plays an important role in investigating the problems of dynamic symmetry breaking and fermion mass generation
in gauge theories. The main question that has been debated for a long time is whether there is a critical fermion flavor number,
$ N_{cr}$, where the separation of the chiral symmetric phase and the phase with broken chiral symmetry occurs
(see  \cite{Appelquist1986,Kotikov1993,Atkinson1990,Gusynin:2016som,Karthik2019}).
Analytical studies of chiral symmetry breaking and mass generation in QED$_3$ are usually based on the use of the Schwinger-Dyson
equations with some ansatzes for the full fermion-photon vertex.

Like QED in 4 dimensions, QED$_3$ has the
important property - the covariance of the fermion propagator
and vertex in the Landau-Khalatnikov-Fradkin (LKF) transformations  [\onlinecite{Landau:1955zz,Johnson:1959zz}].
These transformations have a simple form in representing a coordinate space and allow us to compute Green's functions in an
arbitrary covariant gauge if we know their value at any particular gauge (for application of the LKF transformations,
see papers  ~[\onlinecite{Bashir:2000rv}]
and the review in Ref.~[\onlinecite{Bashir:2007zza}]). 

In the recent paper \cite{Gusynin:2020cra}, we studied the gauge-covariance of the massless fermion propagator in
quenched QED$_3$ in a linear covariant gauge.
We recall here that the quenched limit of QED is the approximation in which we can neglect the effects of closed fermion loops.
This approximation arose in the study of the lattice representation of QED$_4$ (see [\onlinecite{Marinari:1981qf}]), which showed
that a reasonable estimate of the hadron spectrum can be obtained by eliminating all internal quark loops. Moreover, the quenching
approximation in QED$_4$  is now used to include QED effects in lattice QCD calculations (see the recent paper [\onlinecite{Hatton:2020qhk}]
and discussions therein).

Immediately after its introduction in the study of the lattice representation QED$_4$, the quenched approximation in QED$_4$ was also
used in Refs.~[\onlinecite{Fomin:1984tv,Leung:1985sn,Gusynin1999PRD}] within the framework of the formalism based on the study of the Schwinger-Dyson
equations.

In Ref. \cite{Gusynin:2020cra}, following \cite{Gusynin1999PRD,Kotikov:2019bqo,James:2019ctc}
we applied dimensional regularization
and studied  the self-consistency of the LKF transformation of the massless fermion propagator in
quenched QED$_3$ in a linear covariant gauge.
Analysis of \cite{Gusynin:2020cra} led to the conclusion that in exactly three dimensions, $d = 3$, all odd perturbative coefficients, starting from the third order,
must be equal to zero in any gauge if QED$_3$ does not have (infrared) singularities, as discussed in \cite{Jackiw1981,Karthik2017}.

To test this, in Ref. \cite{Pikelner:2020mga} we calculated the three- and four-loop orders and found that the three-loop corrections are finite and gauge-invariant,
while the four-loop corrections have singularities.
The terms depending on the gauge parameter are completely determined by lower orders in accordance with the LKF transformation.

The paper is organized as follows. Section II is devoted to the results of the LKF transformation for the fermionic propagator in momentum space.
Section III presents the results of three- and four-loop corrections to the fermionic propagator and their correspondence to the results of the LKF transformation.
In addition, some predictions above four loops are presented. The results are summarized and discussed in Section IV.
In Appendix A we show some details of the calculations.

\section{LKF transformation}
\label{LKF-d=3}

In what follows, we will consider a Euclidean space of dimension $d=3 - 2\veps$.
The general form of the fermionic propagator $S_F(p,\xi)$ in some gauge $\xi$ is:
\be
S_F(p,\xi) = \frac{\I}{\hat{p}} \, P(p,\xi) \, ,
\label{SFp}
\ee
where the tensor structure is highlighted, i.e. factor $\hat{p}$ containing $\gamma$ Dirac matrices, and $P(p,\xi)$ is a scalar function of $p=\sqrt{ p^2}$.

Following to Ref. \cite{Gusynin:2020cra} the fermion propagator can be represented
in the form
\be
P(p,\xi) = \sum_{m=0}^{\infty} a_m(\xi)\, \left(\frac{\al}{2\sqrt{\pi}\,p}\right)^m\,
{\left(\frac{\overline{\mu}^2}{p^2}\right)}^{m\ep} \, ,
\label{Pxi:QED3}
\ee
where $a_m(\xi)$ are coefficients of the loop expansion of the propagator $P(p,\xi)$ and
$\overline{\mu}$ is the $\overline{MS}$-scale.

It is also convenient to introduce the $x$-spatial representation $S_F(x,\xi)$ of the fermionic propagator in the form:
\be
S_F(x,\xi) =  \hat{x} \, X(x,\xi) \, ,
\label{SFx}
\ee
which is related by the Fourier transform to $S_F(p,\xi)$ in (\ref{SFp}).

The LKF transformation
expresses the covariance of the fermion propagator under a gauge transformation.
It can be derived by standard arguments, see, \eg, [\onlinecite{Landau:1955zz,Johnson:1959zz}]
and its
general form can be written as
(see Refs. \cite{Kotikov:2019bqo,James:2019ctc})
\be
S_F(x,\xi) = S_F(x,\eta)\,e^{D(x)}\, ,~~
%
D(x) = e^2\,\Delta\,\mu^{2\veps}\,\int \frac{\D^d q}{(2\pi)^d } \, \frac{e^{-\I q x}}{q^4},~~ \Delta=\xi -\eta \, ,
\label{def:D(x):QED3}
\ee
in $d=3-2\ep$. Here $\eta$ is some other gauge. 

The calculation yields \cite{Gusynin1999PRD}:
\be D(x) = - \frac{\al\,\Delta}{2\pi\,\mu}\,\frac{\Gamma(1/2-\ep)}{1+2\ep} \,
(\pi \mu^2 x^2)^{1/2+\ep}
\, .
\label{DxQED3}
\ee

The LKF transformation (\ref{def:D(x):QED3}) relates \cite{Gusynin:2020cra} the coefficients $a_k(\xi)$ in Eq. (\ref{Pxi:QED3}) and $a_m(\eta)$ in a similar representation
as
\be
a_k(\xi) =
\sum_{m=0}^{k}
\, (-2 \Delta)^{k-m} \, a_m(\eta) \, \hat{\Phi}(m,k,\ep) \  \phi(k-m,\ep),
\label{am.xid3}
\ee
where
\begin{flalign}
\hat{\Phi}(m,k,\ep)
= \frac{\Gamma(3/2-m/2-(m+1)\ep)\Gamma(1+k/2+k\ep)}{\Gamma(1+m/2+m\ep)
  \Gamma(3/2-k/2-(k+1)\ep)}
\label{hPhi}
\end{flalign}
and
\be
\phi(l,\ep) = \frac{\Gamma^l(1/2-\ep)}{l!\,(1+2\ep)^l\Gamma^l(1+\ep)}\, .
\label{phi}
\ee

Now consider $a_{m}(\xi)$ with $m\leq 4$.
Taking into account the first two orders of the $\ep$-expansion, we have:
\bea
&&a_{0}(\xi)=a_{0}(\eta) =1 \, ,~~
a_{1}(\xi)=a_{1}(\eta) - \frac{\pi}{2} \, \delta \, \Bigl(1+2\ep(l_2-1)\Bigr) \, a_{0}(\eta)\, , \nonumber \\
&&a_{2}(\xi)=a_{2}(\eta) - \frac{4}{\pi} \, \delta \, \Bigl(1-2\ep(l_2+1)\Bigr) \, a_{1}(\eta) + \delta^2 \,\Bigl(1-4\ep\Bigr)
\, a_{0}(\eta)\, , \nonumber \\
&&a_{3}(\xi)=a_{3}(\eta)  + 6\pi\ep \, \delta \, a_{2}(\eta) - 12\ep \, \delta^2 \, a_{1}(\eta)
+2\pi\ep \, \delta^3 \, a_{0}(\eta)\, , \nonumber \\
&&a_{4}(\xi)=a_{4}(\eta) - \frac{2\delta}{3\pi\ep} \, \Bigl(1+2\ep(3-l_2)\Bigr)\, a_{3}(\eta) - 2 \delta^2 \, a_{2}(\eta)
+ \frac{8\delta^3 }{3\pi} \, a_{1}(\eta)
-\frac{\delta^4 }{3} \, a_{0}(\eta)\, ,
\label{axi16}
\eea
where $\delta=\sqrt{\pi}\Delta$ and $l_2=\ln 2$.

Setting $\eta=0$, i.e. choosing the initial Landau gauge, we can represent the results (\ref{axi16}) for $a_m(\xi)$ in the form
\be
a_m(\xi)= a_m(0) + \,\xi \, \tilde{a}_m(\xi)
\label{am=am0}
\ee
and verify that our results for $\tilde{ a}_{m}(\xi)$ are completely determined by $a_{l}(0)$, $(l < m)$, i.e. coefficients of lower orders in accordance with the LKF
transformation.

\section{Fermion propagator: three- and four-loop coefficients
}
\label{Sec:Conclusion}

For the calculations, it is convenient to use
\be
 P(p,\xi) = \frac{1}{1-\sigma(p,\xi)} \, ,
\label{SFp.1}
\ee
where the 1PI-part $\sigma(p,\xi)$ can be represented (similary to (\ref{SFp})) as
\be
\sigma(p,\xi) = \sum_{m=1}^{\infty} \sigma_m(\xi)\, \left(\frac{\al}{2\sqrt{\pi}\,p}\right)^m\,
{\left(\frac{\overline{\mu}^2}{p^2}\right)}^{m\ep} \, .
\label{Sigmaxi:QED3}
\ee

Some details of the calculations will be shown below in Appendix A.
Here we present the results for $\sigma_m(\xi)$, which can be represented in the form similar to (\ref{am=am0}):
\be
\sigma_m(\xi)= \sigma_m(0) + \,\xi \, \tilde{\sigma}_m(\xi)
\, .
\label{sim=sim0}
\ee

Taking into account the first two orders of the $\ep$-expansion, we have for the coefficients
$\sigma_m(0)$
\bea
&&\hspace{-1cm}
\sigma_1(0)=0;~~
\sigma_2(0)=\pi\left[\frac{3\pi^2}{4}-7 -\Bigl((1-3l_2)\pi^2 +12\Bigr)\ep \right],~~ \nonumber \\
&&\hspace{-1cm}
\sigma_3(0)= \pi^{5/2}\left[\frac{43\pi^2}{4}-105 +\ep \left\{ 2 \Bigl(185-105l_2+137\zeta_3\Bigr)
  -\frac{\pi^2}{6} \Bigl(451-171l_2\Bigr)\right\}\right], \nonumber \\
&&\hspace{-1cm}\sigma_4(0)=
\pi^2 \left[ \left(\frac{43}{6}\pi^2-70\right) \, \frac{1}{\ep} +
  \overline{\sigma}_{4} + \frac{5954}{3} + \frac{173}{18}\,\pi^2- \frac{513}{10}\,\pi^4\right]
\, ,
  \label{si4.0} 
  \eea
  where
$\overline{\sigma}_{4}$ contains the most complicated part
\be
\overline{\sigma}_{4} = 209l_2^4+5016a_4+4264{\rm Cl}_4(\pi/2)+ \left(\frac{533}{3}{\rm C} -930\,l_2\right) \, \pi^2
+\frac{2078}{3}\zeta_3 \, .
 \label{oa40}
 \ee
 and
 \be
 a_4={\rm Li}_4(1/2),~~\zeta_n={\rm Li}_n(1)  \, ,
 \label{l2} 
\ee
${\rm C}$ is Catalan, ${\rm Li}_n$ are polylogarithms and ${\rm Cl}_4$ is Clausen number.

With the same accuracy, we have for the coefficients
$\tilde{\sigma}_m(\xi)$
\bea
&&\hspace{-1cm}
\tilde{\sigma}_1(\xi)=-\frac{\pi^{3/2}}{2} \Bigl(1-2(1-l_2)\ep\Bigr),~~
\tilde{\sigma}_2(\xi)=\pi \, \xi \left[1-\frac{\pi^2}{4} -\bigl(4-(1-l_2)\pi^2\bigr)\ep\right],~~ \nonumber \\
&&\hspace{-1cm}\tilde{\sigma}_3(\xi)= \pi^{5/2}\, \Biggl[\frac{3\pi^2}{4}-7 + \left(1-\frac{\pi^2}{8}\right)\xi^2 +
  \ep \Biggl\{-40-14l_2+ \frac{\pi^2}{2} \bigl(4+9l_2\bigr) \nonumber \\
&& \hspace{1cm} + \left(2l_2-4+ \frac{3\pi^2}{4} \bigl(1-l_2\bigr) 
  \right) \xi^2 \Biggr\}
\Biggr] ; \nonumber \\
&&\hspace{-1cm}\tilde{\sigma}_4(\xi)= \pi^{2}
\Biggl[\left(70-\frac{43\pi^2}{6} \right)\, \frac{1}{\ep}
  + \frac{520}{3}- \frac{\pi^2}{9} \Bigl(881+42l_2\Bigr) + \frac{129\pi^4}{27}
  -\frac{548}{3} \zeta_3  \nonumber \\
&&\hspace{1cm}
  + \xi \, \left(28-\frac{33\pi^2}{4} + \frac{9\pi^4}{16} \right)
+ \xi^3 \, \left(-\frac{4}{3} +\frac{3\pi^2}{4}- \frac{\pi^4}{16} \right)\Biggr]\, .
\label{tsi4xi}
\eea

Note that the finite parts of the coefficients $\sigma_1(\xi)$ and $\sigma_2(\xi)$ coincide with the corresponding ones in Ref. \cite{Kotikov:2013eha}.
So, we see that
\be
\sigma_{4}(\xi)
= \pi^2 \left(\frac{43}{6}\pi^2-70\right) \,  \frac{(1-\xi)}{\ep} + O(\ep^0) \, ,
\label{si4.0A} 
  \ee
  i.e. the four-loop results are finite in Feynman gauge.

  \subsection{$a_m(\xi)$}
\label{Sec:Conclusion}

 The coefficients  $a_m(\xi)$ and $\sigma_m(\xi)$ are related each other as
\be
a_1=\sigma_1,~~a_2=\sigma_2+\sigma_1^2,~~a_3=\sigma_3+2\sigma_2\sigma_1+\sigma_1^3,~~
a_4=\sigma_4+2\sigma_3\sigma_1+\sigma_2^2+3\sigma_2\sigma_1^2+\sigma_1^4 \, .
\label{a.sigma}
\ee

Since $\sigma_1(\xi) \sim \xi$, we see in (\ref{a.sigma}) that $a_i(0)=\sigma_i(0)$ for $i \leq 3$ and thus so $a_i (0)$ with $i \leq 3$ can be found
in Eq. (\ref{si4.0}). According to (\ref{a.sigma}) we have for $a_{4}(0)$:
\be
\hspace{-1cm}a_{4}(0) = \sigma_4(0)+ \pi^2 {\left(\frac{3\pi^2}{4}-7\right)}^2=
\pi^2 \left[ \left(\frac{43}{6}\pi^2-70\right) \, \frac{1}{\ep} +
  \overline{\sigma}_{4} + \frac{6101}{3} - \frac{8}{9}\,\pi^2- \frac{4059}{80}\,\pi^4\right]\, .
\label{a4.0A} 
\ee
 
With the same accuracy, we have for the coefficients  $\tilde{a}_m(\xi)$
\bea
&&
\hspace{-1cm}
\tilde{a}_1(\xi)=\tilde{\sigma}_1(\xi)=-\frac{\pi^{3/2}}{2} \Bigl(1-2(1-l_2)\ep\Bigr),~~
\tilde{a}_2(\xi)=\pi \, \xi \Bigl(1-4\ep\Bigr),~ \nonumber \\
&&\hspace{-1cm}
\tilde{a}_3(\xi)= \pi^{5/2}\,\ep \left(\frac{43\pi^2}{4}-105 +2\xi^2\right), \nonumber \\
&&\hspace{-1cm}\tilde{a}_4(\xi)= \frac{\pi^{2}}{3}\left[\left(210-\frac{43\pi^2}{2} \right)\, \frac{1}{\ep}
  + 520+ \frac{2\pi^2}{3}\Bigl(32-21l_2\Bigr) -548\zeta_3 + 6\xi \, \left(7-\frac{3\pi^2}{4}\right)-\xi^3\right]
\, .
\label{ta4xi}
\eea

Note that the finite parts of the coefficients $a_1(\xi)$ and $a_2(\xi)$ coincide with the corresponding ones in
[\onlinecite{Bashir:2000rv}] (see also the Ref. \cite{Gusynin:2020cra} and discussions therein).

We see that the coefficients $\tilde{a}_m(\xi)$ $(m=2,3,4)$ have a simpler form than the corresponding coefficients
$\tilde{\sigma}_m(\xi)$. Moreover, as in the case of $\sigma_{4}(\xi)$, we also see that
\be
a_{4}(\xi) = \sigma_{4}(\xi)  + O(\ep^0) =
\pi^2 \left(\frac{43}{6}\pi^2-70\right) \, (1-\xi) \, \frac{1}{\ep} + O(\ep^0) \, ,
\label{a4.0B} 
  \ee
  i.e. the four-loop results are finite in Feynman gauge.

\subsection{Beyond four-loops}
\label{LKF-d=3}

What can we say about higher orders using the LKF transformation?

Consider $a_{5}(\xi)$ and $a_{6}(\xi)$. From Ref. \cite{Gusynin:2020cra} we have
\bea
&&a_{5}(\xi)=a_{5}(\eta)  + \frac{45}{2}\pi\ep \, \delta \, a_{4}(\eta) - \frac{15}{2} \, \delta^2 \, a_{3}(\eta)
\nonum \\
&&\qquad ~~~\,-15\pi \ep \, \delta^3 \, a_{2}(\eta) + 15\ep \, \delta^4 \, a_{1}(\eta)
- \frac{3}{2} \pi\ep \, \delta^5 \, a_{0}(\eta)\, , \nonumber \\
  &&a_{6}(\xi)=a_{6}(\eta) + \frac{4\delta}{5\pi\ep} \, a_{5}(\eta) - 9 \delta^2 \, a_{4}(\eta)
+ \frac{2\delta^3 }{\pi\ep} \, a_{3}(\eta)
\nonum \\
&&\qquad ~~~\,+3 \delta^4 \, a_{2}(\eta)
- \frac{12\delta^5 }{5\pi} \, a_{1}(\eta) + \frac{\delta^6 }{5} \, a_{0}(\eta)\, .
\label{axi56}
\eea

Now we take the $\eta$-gauge as the Feynman gauge (since $a_{4}(\xi=1)$ is finite)
and consider $a_{5}(\xi)$ and $a_{6}(\xi)$ with accuracy $O(\ep)$ and $O(\ep^0)$ respectively. So we have
\bea
&&a_{5}(\xi)=a_{5}(1)  - \frac{15}{2} \, \pi \, (\xi-1)^2 \, a_{3} + O(\ep)\, , \nonumber \\
&&a_{6}(\xi)=a_{6}(1) + \frac{4(\xi-1)}{5\sqrt{\pi}\ep} \, a_{5}(1) 
+ \frac{2\sqrt{\pi}(\xi-1)^3 }{\ep} \, a_{3} + O(\ep^0)\, ,
\label{axi56a}
\eea
where we take into account that the finite part of $a_{3}$ is gauge-independent.

Thus, we see that the LKF transformation gives some information about the $\xi$-dependence of $a_{5}(\xi)$ and $a_{6}(\xi)$, but to obtain exact results,
it is necessary to evaluate these values in a particular gauge.

\section{
  Conclusion}
\label{Sec:Conclusion}

In our recent paper \cite{Gusynin:2020cra} we studied the LKF transformation for the QED$_3$ massless fermionic propagator in the quenched approximation.
  Studying this transformation in dimensional regularization, we found that the contributions of odd orders, starting from the third, to even ones,
  are accompanied by singularities that look like $\ep^{-1}$ in dimensional regularization.
  In turn, the even orders produce contributions to the odd ones, starting from the third, $\sim \ep$.

  Following the arguments in favor of the perturbative finiteness of the massless quenched QED$_3$ \cite{Jackiw1981,Karthik2017} and assuming
  the existence of a finite limit at $\ep \to 0$, in \cite{Gusynin:2020cra} we have shown exactly in $d=3$ that {\it all odd terms of $a_{2t+1}(\xi)$
    in perturbation theory except $a_{1}$ must be exactly zero in any gauge.}

  This statement was very strong and needed verification, which was done exactly in Ref. \cite{Pikelner:2020mga}.
  We calculated three- and four-loop corrections, i.e. terms $a_{3}(\xi)$ and $a_{4}(\xi)$, directly in the framework of perturbation theory.
  We found that $a_{3}(\xi)$ is finite and gauge independent when $\ep \to 0$.
  The coefficient $a_{4}(\xi)$ has singularities, which violates the status of the infrared perturbative finiteness of the massless quenched QED$_3$.
  The obtained results have the following property: all terms that depend on the gauge parameter are completely determined by lower orders in accordance
  with the LKF transformation.

  Moreover, in Ref. \cite{Pikelner:2020mga} we found that the singularities contributing to the coefficient $a_{4}(\xi)$, $\sim (1-\xi)$ and thus $a_{ 4}(\xi)$ 
  is finite in the Feynman gauge. The reason for this effect is not clear and more research is needed to elucidate it.

\acknowledgments
Author thanks the Organizing Committee of the International Conference ``Models in Quantum Field Theory''.
for the invitation.\\

{\bf Conflict of Interest}: The authors declare that they have no conflicts of interest.

\appendix
\def\theequation{A\arabic{equation}}
\setcounter{equation}{0}

\section{Details of calculations}
\label{sec:calc}
To calculate the unrenomalized fermion self-energy in QED$_3$ up to the four-loop order, authors of \cite{Pikelner:2020mga} used the QCD result for the
unnormalized quark propagator. The QCD result valid for an arbitrary space-time dimension $d$ and for an arbitrary gauge in the form of a final
answer reduced to the set of master integrals is available in \cite{Ruijl:2017eht}, and also shipped with the \texttt{FORCER} package\cite{Ruijl:2017cxj},
designed to reduce the four-loop integrals of the massless propagator. The
QED$_d$ limit from the QCD result was obtained by substituting:
\begin{equation}
  \label{eq:qed2qcdSub}
  C_A = d_A^{abcd}d_A^{abcd} = d_A^{abcd}d_F^{abcd} =0, \quad
  C_F = d_F^{abcd}d_F^{abcd} = T_F=1\,.
\end{equation}
After that, the quenched QED$_d$ limit is obtained by setting $n_f = 0$, which discards all diagrams with closed fermionic loops.

To calculate the required four-loop propagator integrals in $d=3-2\varepsilon$, the dimensional recurrence and analyticity (DRA) method \cite{Lee:2009dh}
was used,
giving results in the form of fast convergent sums. After performing summation, high-precision numerical values for integrals in an arbitrary
space-time dimension can be reconstructed to obtain analytical answers using the \texttt{PSLQ} algorithm \cite{PSLQ}  after determining an
adequate basis of transcendental constants.
 
Note that near $d = 4$ such calculations give expansions of all necessary master integrals \cite{Lee:2011jt}. The results are well known and
are available in the input form for the \texttt{SummerTime} package \cite{Lee:2015eva} along with the package itself, as well as from
\cite{Magerya:2019cvz}.

The case $d=3-2\ep$ is less known and was recently considered in \cite{Lee:2015eva}, from which the $\ep$-expansion of most of the necessary master
integrals for our calculation is available. Successful reconstructions \cite{Lee:2015eva} around $d = 3$ have been performed using a basis of
transcendental constants consisting of only multiple zeta values (MZV) and alternating MZVs. As already noted by the authors of \cite{Lee:2015eva},
such a basis is too restrictive to represent all master integrals, so some of them remained unreconstructed.

In Ref. \cite{Pikelner:2020mga}, all necessary integrals were successfully restored and agreement was found with the results of \cite{Lee:2011jt}
using a basis consisting of MZVs and alternating MZVs.  In addition, one more constant was determined, which was unknown in \cite{Lee:2015eva} after
careful analysis of one of the integrals, namely, $D_1(p)$:
\be
  D_1(p)=\int \, \frac{d^dk_1d^dk_2}{(2\pi)^{2d}} \,
  \frac{L(k_2)L(p-k_1)}{k_1^{2}(p-k_2)^{2}(k_1-k_2)^{2}} \, ,
  \label{D1def}
  \ee
  where $L(p)$ is simple loop:
  \be
  L(p)=\int \, \frac{d^dk}{(2\pi)^{d}} \,
  \frac{1}{k^{2}(p-k)^{2}} \, .
  \label{Ldef}
  \ee

  Evaluating loops, we have that
  \be
  D_1 = \frac{1}{(4\pi)^{d}} \, \pi^3 \,G(p;1,1/2,1,1/2,1) + O(\ep) \, ,
  \label{D1}
  \ee
  where
  \be
  G(p;\alpha_1,\alpha_2,\alpha_3,\alpha_4,\alpha_5)=\int \, \frac{d^dk_1d^dk_2}{(2\pi)^{2d}} \,
  \frac{1}{k_1^{2\alpha_1}k_2^{2\alpha_2}(p-k_2)^{2\alpha_3}(p-k_1)^{2\alpha_4}(k_1-k_2)^{2\alpha_5}} \, .
  \label{G12345}
  \ee

  The diagram $G(p;\alpha,1,\beta,1,1)$ was studied in \cite{Kotikov:2013eha}, where several of its representations were presented as combinations of
  ${}_3F_2$-hypergeometric functions with argument 1.
\footnote{This analysis is based on \cite{Kotikov:1995cw}, where a class of more complicated diagrams with three arbitrary indices was studied and the
  corresponding results were also given as combinations of ${}_3F_2$-hypergeometric functions with argument 1.}
Considering representations in the case $\alpha=\beta=1/2$, Andrey Pikelner noticed that only  generalized polylogarithms (GPLs) appear with an argument
in the form of the fourth root of unity. By extending \texttt{PSLQ} framework to include the full GPL set with fourth root of unity, he
  successfully reconstructed the analytic result for $G(p;1,1/2,1,1/2,1)$ as
  \be
  G(p;1,1/2,1,1/2,1) = \frac{1}{(4\pi)^{d}} \, \frac{8}{3\pi} \Bigl( \mathcal{C} \pi^2 + 24 \mathrm{Cl}_4\left(\frac{\pi}{2}\right)+  O(\ep^1)\Bigr)  \,
  \frac{\mu^{2\ep}}{p^{2(1+2\ep)}}  \, .
  \label{G12345.1}
  \ee
  Here $ \mathcal{C}= \mathrm{Cl}_2(\pi/2)$ is the Catalan constant and $\mathrm{Cl}_n(\theta)$ is the Clausen function, which for an even
  weight can be expressed in terms of the
  classical polylogarithm as $\mathrm{Cl}_{2k}(\theta) = {\rm Im} {\rm Li}_{2k} (e^{i\theta})$.
  As can be understood from the above result, the required extension of the basis of
  transcendental constants includes polylogarithms with a fourth root of unity, in this case the Clausen function (see, for example, \cite{Laporta:2018eos},
  where GPLs appear as arguments with the second, fourth, and sixth roots of units).

  Such a diagram is useful in calculations within effective theories. Moreover, following the transformations discussed in \cite{Kotikov:1995cw}, we see
  that for $d=3$
  \bea
  &&G(p=1;1,1/2,1,1/2,1)=G(p=1;1/2,1,1/2,1,1/2) \nonumber \\
  &&=G(p=1;1,1/2,1/2,1/2,1)=G(p=1;1/2,1,1,1,1/2) \, 
   \label{G12345.2}
   \eea
and thus the result (\ref{G12345.1}) is already applicable for several diagrams.


\end{document}